\documentclass{llncs}
\usepackage{makeidx}  % allows for indexgeneration
\usepackage{amsmath, amssymb}
\usepackage{textcomp}
\usepackage{graphicx}
\usepackage{subfig}
\usepackage{multirow}
\begin{document}

\title{An Auction Mechanism for Resource Allocation in Mobile Cloud Computing Systems}
\titlerunning{Auction MCC}  % abbreviated title (for running head)
\author{Yang Zhang \and Dusit Niyato \and Ping Wang}
\authorrunning{Yang Zhang et al.} % abbreviated author list (for running head)
\tocauthor{Dusit Niyato, and Ping Wang}
\institute{School of Computer Engineering, Nanyang Technological University (NTU), Singapore}

\maketitle              % typeset the title of the contribution

\begin{abstract}
A mobile cloud computing system is composed of heterogeneous services and resources to be allocated by the cloud service provider to mobile cloud users. On one hand, some of these resources are substitutable (e.g., users can use storage from different places) that they have similar functions to the users. On the other hand, some resources are complementary that the user will need them as a bundle (e.g., users need both wireless connection and storage for online photo posting). In this paper, we first model the resource allocation process of a mobile cloud computing system as an auction mechanism with premium and discount factors. The premium and discount factors indicate complementary and substitutable relations among cloud resources provided by the service provider. Then, we analyze the individual rationality and incentive compatibility (truthfulness) properties of the users in the proposed auction mechanism. The optimal solutions of the resource allocation and cost charging schemes in the auction mechanism is discussed afterwards. 
\end{abstract}

\section{Introduction}
In a cloud computing system, the service provider has different resources to be leased or sold to cloud users. The users use the allocated resources to run their applications. A mobile cloud computing system differs from general cloud computing systems in some aspects. One important aspect is the combination pattern of demanded services. In a general cloud application, a cloud user may request either a single service or a combination of services. For example, a user of Amazon's EC2~\cite{Amazon.EC2.Web} may subscribe for a server only, or spend extra cost to subscribe for other resources and services only when necessary. However, services in a mobile cloud are generally provided in bundles. The reason is that a mobile device as an end user is relatively a ``thin client'' which cannot process too complex tasks. Therefore, most of the tasks are offloaded to the cloud side, and the cloud service providers need to provide a bundle of service to process the tasks. For example, in a mobile game service, the user will need both a processing resource for artificial intelligent player and a storage for game module. Most importantly, communication bandwidth, as crucial resource in mobile systems, should be guaranteed for transferring data of mobile cloud applications.

\begin{figure*}[!t]
		\begin{center}
		\subfloat[]{\includegraphics[scale=0.34]{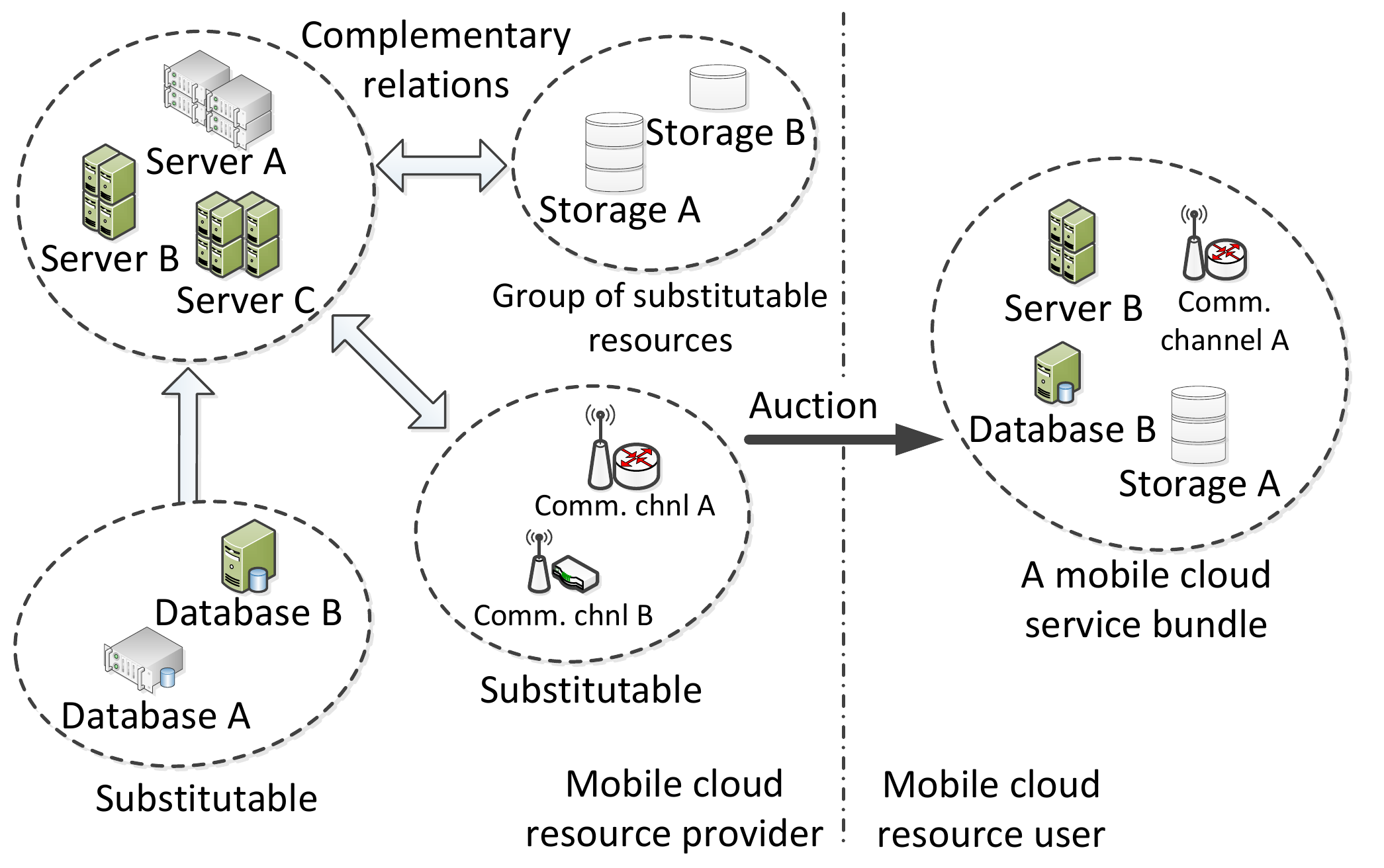}}
		\subfloat[]{\includegraphics[scale=0.39]{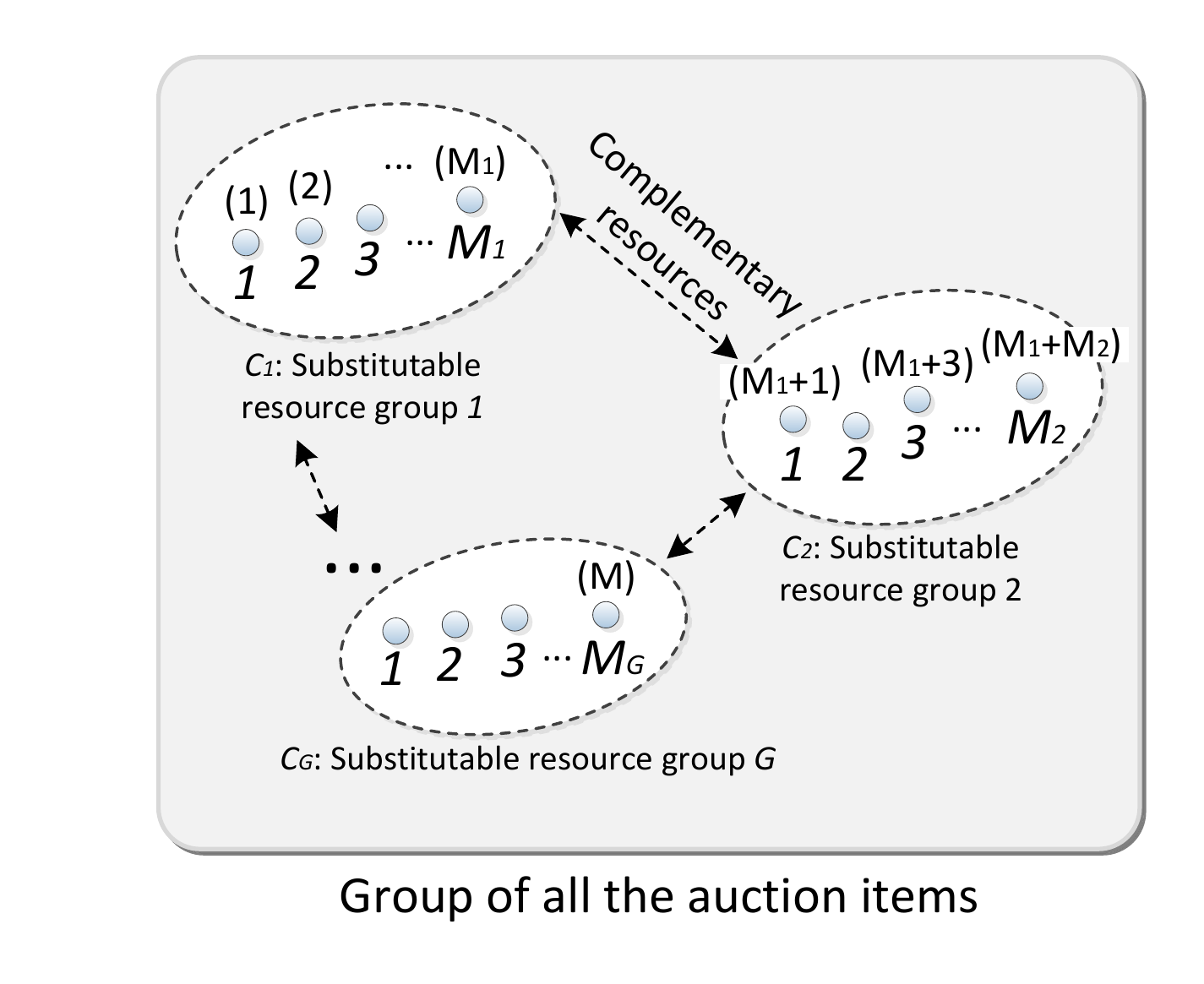}}
		\end{center}
		\caption{\small Asymmetric scenario: (a) user 1's utility to her type $t_1$, (b) user 2's utility to user 1's type $t_1$, and (c) the service provider's revenue to user 1's type $t_1$. Premium and discount factors' impacts: (d) user's utility under different settings of premium and discount factors.}
		\label{fig:SystemModel}
		\end{figure*}

In this paper, we model the resource allocation problem in a mobile cloud system as a combinatorial auction with substitutable or complementary commodities. The service provider is defined as a seller, while the users are defined as buyers. The available cloud resources are to be sold and allocated by the seller to buyers. Moreover, we design the proposed auction mechanism for the mobile cloud computing system, which is proved to be individual rational and incentive compatible (as shown in Section~\ref{tag:subsec.iric}). As shown in Fig.~\ref{fig:SystemModel}(a), the cloud resources of the service provider can be categorized into several groups, e.g., processing (i.e., server), storage, and communication resources. Resources in the same group are different in quality, but have similar functions. These heterogeneous resources can be allocated in bundles. Therefore, we consider the resources as commodities which could be substitutable or complementary. If the resources are substitutable, the valuations of these resources are sub-additive (i.e., total valuation of all resources obtained altogether is less than the sum of valuation of each resource). For example, any server can be treated by the user as a substitutable resource. On the other hand, if the resources are complementary (e.g., a server for processing and bandwidth for data transmission where the user needs both), the valuations of these resources are super-additive (i.e., total valuation of resources obtained altogether is higher than the sum of valuation of each resource). 

The rest of this paper is organized as follows. In Section~\ref{tag:sec.relatedwork}, we review related work. We describe the system model in Section~\ref{tag:sec.sysmodel}. In Section~\ref{tag:sec.analysis}, we discuss the individual rationality and the incentive compatibility properties of cloud users. Also, the optimal solutions of the service provider's resource allocation and cost charging schemes are presented. Section~\ref{tag:sec.numerical} shows the numerical results. Finally, we conclude and summarize the paper in Section~\ref{tag:sec.conclusion}.

		%

%--------------------------------------------------------------------
\section{Related Work} \label{tag:sec.relatedwork}
In mobile cloud computing, the tasks of mobile applications will be partly offloaded to servers in cloud. Therefore, the performance of mobile applications can be improved (e.g. computer games~\cite{z.li.cose.2011}). A few works addressed the issues of general/mobile cloud computing systems. In~\cite{Armbrust.Cloud.2010}, the definition and discussion of general cloud computing systems were presented. \cite{Wang.MCCthesis.2011} focused on cloud computing implemented on mobile network infrastructures, i.e., mobile cloud computing systems. Also, \cite{Hoang.Survey.2011} presented a comprehensive survey of mobile cloud computing including the system architecture, applications, resource allocation and other issues.

Auction is a general and effective way for resource allocation. A tutorial introduction of auction theory for computer science was presented in~\cite{Parsons.Tut.2011}. Specifically, \cite{Myerson.OAD.1981} developed an auction mechanism for a single auction commodity scenario, and proposed the concepts of individual rationality and incentive compatibility in auction mechanism designs. \cite{Branco.MUAI.1996} extended the model in \cite{Myerson.OAD.1981} to a multiple commodity auction scenario. However the auction commodities in \cite{Branco.MUAI.1996} are independent with each other and indivisible. \cite{Levin.OAC.1997} discussed an auction model with two complementary auction commodities. \cite{Burguet.CPAS.2005} explored the auction designs for substitutable commodities. \cite{Mullins.Patent.2010} proposed an auction technique to optimize cloud resource distribution in a cloud computing system. \cite{Lin.DAMC.2010} proposed a second-price auction mechanism to allocate a single type of cloud resource, i.e., computational capacity.

To our best knowledge, the analytical auction model for a mobile cloud computing system containing both complementary and substitutable resource was not considered before.

%--------------------------------------------------------------------
\section{System Model and Assumptions} \label{tag:sec.sysmodel}
In this section, we first present the description of the mobile cloud computing system and auction mechanism. Then, the utility of users is defined.
	\subsection{Mobile Cloud Computing System and Auction Mechanism}	\label{sec:subsec.sysmodel.auction}	
	We consider a mobile cloud computing system with a service provider offering mobile cloud applications/services to $N$ users (Fig.~\ref{fig:SystemModel}(a)). The service provider has cloud resources (e.g., CPU/computational capacity, database, and communications) to support the offered services. There are totally $M$ resources which are divided into $G$ groups. Group $k$ contains $M_k$ resources, for $k=1,2,\ldots,G$. In the same group, the resources provide similar functions, and hence the resources are substitutable. On the other hand, in different groups, the resources provide different functions, and hence the resources are complementary in building cloud services for the users.
	
	The substitutable and complementary resources in a mobile cloud computing system is shown in Fig.~\ref{fig:SystemModel}(b) which is an abstracted model of Fig.~\ref{fig:SystemModel}(a). We assume that the transitivity condition of the substitutable and complementary resources holds, i.e.,
	\begin{itemize}
		\item Resources A, B and C are substitutable/complementary mutually, if and only if, resources A and B, as well as B and C are substitutable/complementary.
		\item Resources A and C are complementary, if resources A and B are complementary, while resources B and C are substitutable.
	\end{itemize}
	By assuming the transitivity condition, there is no ``triangle'' relations among any three resources A, B and C such that A and B are substitutable (in one group), B and C are complementary (in different groups), but A and C are still substitutable.
	%Consequently, a substitutable group and a complementary group will not intersect.
	
The auction mechanism is developed for resource allocation by the mobile cloud service provider to the resource buyers (i.e., users). The auction mechanism works as follows. First, the users submit bids containing users' valuations on the resources to the service provider. The valuation of user $i$ on resource $j$ is denoted by $v_{ij}$. There are other users who also compete for the resources. Therefore, the valuation depends not only on the type (i.e., preference) of user $i$, but also on other users' types. The user $i$'s type, denoted by $t_i$, is defined as the user's private appetite on obtaining the resources. Therefore, the valuation is defined as $v_{ij}(t_i)$, where $\mathbf{t}_{-i}$ is the vector of other users' types except that of user $i$ and $\mathbf{t}=(t_i)=(t_1,t_2,\ldots,t_N)$. The service provider receives the valuation but does not know the user's real type. After receiving the valuations for the resources from all users, the service provider then optimizes the resource allocation to maximize the revenue. The service provider sends back the allocation result denoted by $(p,c)$, where $p$ is the set of all allocation $p_{ij}$, and $c=\{c_1,c_2,\ldots,c_N\}$ is the set of the costs charged to users. $p_{ij}$ is the proportion of resource $j$ allocated to user $i$. $c_i(t_i)$ denotes the cost charged by the service provider to user $i$. Again, the cost is ultimately a function of user's type. Formally, we define $(p,c)$ as a {\em mechanism} revealed by the service provider to all users.

	\subsection{User's Utility} \label{tag:subsec.utility}
	A user has a satisfaction on the allocated resources, which is referred to as the utility. The expected utility of user $i$ to obtain the resources from a group $k$ of substitutable resources can be defined as follows:
	\begin{equation} \label{F_util_in_group}
		u_{i}^{k,s}(p_{i\cdot}^k,c_i^k,t_i) = {\mathbb{E}_{\mathbf{t}_{-i}}} \Big\{ \sum\limits_{r=1}^{M_k} p_{ir}^{k}(t_i) v_{ir}^{k}(t_i) - l_i(t_i)\sum\limits_{r=1}^{M_k}p_{ir}^k(t_i) \Big\} - c_i^k(t_i)
	\end{equation}
	where $t_i$ is the type of user $i$, and $t_i\in[\underline{t_i},\overline{t_i}]$]. $\underline{t_i}$ and $\overline{t_i}$ are the lower bound and upper bound of $t_i$, respectively. In this case, we assume that the type of a user is a random variable with cumulative distribution function (CDF) denoted by $F(t_i)$ and probability density function (PDF) denoted by $f(t_i)$. CDF and PDF are known knowledge or observations to the service provider. $p_{ir}^{k}$ denotes the proportion of $r$th resource in group $k$ allocated to user $i$. $p_{i\cdot}^k$ is $p_{ir}^{k}$ for all $r$. $v_{ir}^{k}$ is user $i$'s private valuation function on the $r$th resource of group $k$. We assume that the valuation functions are positive, non-decreasing and concave with respect to $t_i$. $c_i^k$ is the total cost function of all the resources in group $k$ for user $i$. This cost is paid by user $i$ to the service provider. Note that $l_i(t_i)\sum\nolimits_{r=1}^{M_k}p_{ir}^k(t_i)$ in (\ref{F_util_in_group}) represents the sub-additive term of the substitutable resources. The sub-additive term indicates the loss of utility of a user when multiple resources are allocated from the same group of substitutable resources. Here $l_i(t_i)$ is defined as a discount factor. For example, if the discount factor is large, the user has more dissatisfaction on obtaining the resources from the same group.

	Next, we consider the complementary resources and their contributions in the utility function. First, we define the amount of aggregated resources allocated from group $k$ from the service provider as follows:
		\begin{equation} \label{fml:ProbComp}
			p^{k,c}=\sum\limits_{r=1}^{M_k}p_{ir}^k(t_i).
		\end{equation}
		
	Then, the total utility of user $i$ from obtaining resources can be expressed as follows:
		\begin{equation} \label{fml:TotalUtil}
			u_i(t_i) = {\mathbb{E}_{\mathbf{t}_{-i}}}\Big\{ \sum\limits_{k=1}^{G} u_{i}^{k,s}(p_{i\cdot}^k,c_i^k,t_i) + h_i(t_i)\prod_{k=1}^{G}p^{k,c}(t_i) \Big\}
		\end{equation}
	where $h_i(t_i) \prod_{k=1}^{G}p^{k,c}(t_i)$ defined in (\ref{fml:TotalUtil}) represents the super-additive term of the complementary resources. $h_i(t_i)$ is defined as a premium factor. If the premium factor is large, the user has more satisfaction on obtaining the resources from the different groups. Note that if the user has received nothing from merely one group, the super-additive term will become zero, as a punishment of not receiving a complete bundle of complementary resources required for a mobile cloud computing service.
	
	We will derive the total utility of all $M$ resources next. We substitute (\ref{F_util_in_group}) and (\ref{fml:ProbComp}) into (\ref{fml:TotalUtil}). In this case, the variables with $k$ are mapped to the new variables without $k$. The process is shown in Fig.~\ref{fig:SystemModel}(b). That is, the $r$th resource in group $k$ is mapped to the $j$th resource regardless of the group, where $j=\sum_{i=1}^{k-1}M_k+r$. After mapping, $p_{ir}^{k}$, $v_{ir}^{k}$ and $c_i^k$ become $p_{ij}$, $v_{ij}$ and $c_i$, respectively. $p_{ij}$, $v_{ij}$ and $c_i$ have the same definitions as those in Section~\ref{sec:subsec.sysmodel.auction}. Then we can express the utility $u_i(p_{i\cdot},c_i,t_i)$ of the user $i$ as follows:\footnote{The notation $u_i(t_i)$ is used for short}
		\begin{equation} \label{fml:RealTypeUtil}
			u_i(t_i) = {\mathbb{E}_{\mathbf{t}_{-i}}}\Big\{ \sum\limits_{j=1}^{M}p_{ij}(t_i)v_{ij}(t_i) + S_i(t_i) \Big\} - c_i(t_i)
		\end{equation}
	where $S_i(t_i,{\mathbf{t}}_{-i})$ includes the premium and discount terms defined as follows:
		\begin{equation} \label{fml:STerm}
			S_i(t_i) = h_i(t_i)\prod\limits_{k=1}^{G}\sum_{j=\sum_{n=1}^{k-1}M_k+1}^{\sum_{n=1}^{k}M_n}p_{ij}(t_i) - {l_i(t_i)}\sum_{j=1}^{M}p_{ij}(t_i).
		\end{equation}
	
	We assume in our model that the valuation and extra premium and discount information $S_i(t_i,{\mathbf{t}}_{-i})$ are sent to the service provider. Therefore, the first term in (\ref{fml:RealTypeUtil}) could be treated as user $i$'s ``virtual valuation'' with premium and discount factors. Note, however, that the valuations submitted by users to the service provider may not be the true valuations on the resources. For example, to achieve a higher profit, user $i$ who has the valuation $v_{ij}(t_i)$ on resource $j$ might actually submit a falsified valuation $v_{ij}(\hat{t}_i)$ to the service provider as if the user had the fake type $\hat{t}_i$. Only in the auction mechanism $(p,c)$ that guarantees truthfulness, the users will choose to submit their true valuations as the best bidding strategies. By sending an untruthful valuation, the user $i$'s utility is
	\begin{equation} \label{fml:FakeTypeUtil}
		\hat{u}_i(\hat{t}_i|t_i) = {\mathbb{E}_{\mathbf{t}_{-i}}}\Big\{ \sum\limits_{j=1}^{M}p_{ij}(\hat{t}_i)v_{ij}(t_i) + S_i(\hat{t}_i) \Big\} - c_i(\hat{t}_i).
	\end{equation}

%--------------------------------------------------------------------
\section{Analysis of the Auction Mechanism for Mobile Cloud Computing System} \label{tag:sec.analysis}

In this section, the individual rationality and the incentive compatibility properties of the auction mechanism are analyzed. The optimization of the revenue of the service provider is proposed. Also, the structure of the utility function is discussed.

	\subsection{Individual Rationality and Incentive Compatibility} \label{tag:subsec.iric}
	When an auction and resource allocation mechanism is designed, the auctioneer must ensure positive payoffs of the auction participants (i.e., buyers), so that those buyers are willing to join the auction market. On the other hand, the mechanism should discourage the buyers to submit valuations which are not based on the buyers' true valuation. The individual rationality and incentive compatibility (truthfulness) properties are defined as follows: 
		\begin{equation} \label{fml:Feasible}
			\begin{aligned}
			\text{Individual~rationality}: & \qquad u_i(t_i) \geq 0 \\
			\text{Incentive~compatibility}: & \qquad u_i(t_i) \geq \hat{u}_i(\hat{t}_i|t_i).
			\end{aligned}
		\end{equation}

	\begin{proposition} \label{thm:Prop.ICIR}
	In the proposed auction for the substitutable and complementary resources in a mobile cloud computing system, a mechanism $(p,c)$ is individually rational and incentive compatible if and only if the following conditions hold,
		\begin{equation} \label{fml:Rational}
			u_i(\underline{t_i})\geq0
		\end{equation}
		\begin{equation} \label{fml:IncentiveUIntg}
			u_i(t_i) = u_i(\underline{t_i}) + {\mathbb{E}_{\mathbf{t}_{-i}}}\Big\{ \sum\limits_{j=1}^{M}\int\nolimits_{\underline{t_i}}^{t_i}\frac{\partial{v_{ij}(x)}}{\partial{t_i}}p_{ij}(x)dx \Big\}
		\end{equation}
		\begin{equation}
		\label{fml:IncentiveFinalCond}
			{\mathbb{E}_{\mathbf{t}_{-i}}}\Big\{ \sum\limits_{j=1}^{M}\int\limits_{\hat{t}_i}^{t_i}\frac{\partial{v_{ij}(x)}}{\partial{t_i}}p_{ij}(x)dx \Big\} \geq {\mathbb{E}_{\mathbf{t}_{-i}}}\Big\{ \sum\limits_{j=1}^{M}\big{(}v_{ij}(t_i)-v_{ij}(\hat{t}_i)\big{)}p_{ij}(\hat{t}_i) \Big\}. \\
		\end{equation}
	\end{proposition}
	
	The proof of Proposition~\ref{thm:Prop.ICIR} is provided in the extended paper~\cite{Yang.LongVersion.2013}.

	%---------------------------------------
	%---------------------------------------
	\subsection{The Seller-side Problems: Revenue Maximization and Cost Charging}
	The objective of the service provider is to sell and allocate available resources to the users such that the revenue is maximized. The revenue of the service provider is the sum of the cost $c_i(t_i)$ that each user pays for the allocated resources. The mechanism (i.e., resource allocation) is designed to maximize the total revenue of the service provider, i.e. $\sum_{i=1}^{N} {\mathbb{E}_{t_i}} \{ c_i(t_i) \}$. The following proposition states the optimal mechanism (i.e., maximizing the revenue of the service provider). The following proposition is proved in \cite{Yang.LongVersion.2013}.

	\begin{proposition} \label{thm:Prop.Seller}
	$(p^{*},c^{*})$ is an optimal mechanism if $p^{*}$ maximizes 
		\begin{equation}
		\label{fml:SellerMaxObj}
			\sum\limits_{i=1}^{N}{\mathbb{E}_{\mathbf{t}}} \Big\{ \sum\limits_{j=1}^{M} \big( v_{ij}(t_i) - \frac{1-F(t_i)}{f(t_i)}\frac{\partial{v_{ij}(t_i)}}{\partial{t_i}} \big)p_{ij}(t_i) + S_i(t_i) \Big\}
		\end{equation}
	subject to the constraints in (\ref{fml:Rational}), (\ref{fml:IncentiveUIntg}) and (\ref{fml:IncentiveFinalCond}).
		
	To achieve the optimal revenue, the (maximum) cost $c_i^*(t_i)$ charged to each user $i$ can be obtained from
		\begin{equation}
		\label{fml:SellerOptCost}
			c_i^*(t_i) = {\mathbb{E}_\mathbf{t}} \Big\{ \sum\limits_{j=1}^{M}p_{ij}^*(t_i)v_{ij}(t_i) - \sum\limits_{j=1}^{M}\int\limits_{\underline{t_i}}^{t_i}\frac{\partial{v_{ij}(x)}}{\partial{t_i}}p_{ij}^*(x)dx + S_i^*(t_i) \Big\}.
		\end{equation}
	\end{proposition}
				
	Note that the cost charging and revenue optimization as in Proposition~\ref{thm:Prop.Seller} are essentially linear programming problems, which can be solved numerically by the service provider (seller), provided that the seller has the knowledge of CDF and PDF of buyers' types as we assumed in Section~\ref{tag:subsec.utility}.
	
	%---------------------------------------
	%---------------------------------------
	\subsection{Structures of Utility and Revenue Functions}
	We analyze the structures of the users' utility and the service provider's optimal revenue functions. According to~(\ref{fml:IncentiveUIntg}), each user's utility is only decided by the ``basic utility'' (i.e.,  $u_i(\underline{t_i})$), and the ``marginal utility'' (i.e., ${\mathbb{E}_{\mathbf{t}_{-i}}}\big\{ \sum_{j=1}^{M}\int\nolimits_{\underline{t_i}}^{t_i}\frac{\partial{v_{ij}(x)}}{\partial{t_i}}p_{ij}(x)dx \big\}$).
	
	In the optimization process of the provider's total revenue, as justified in the proof of Proposition~\ref{thm:Prop.Seller} (\cite{Yang.LongVersion.2013}), the basic utility is minimized to be zero by the service provider's resource allocation. Therefore, each user's utility depends solely on the marginal utility.

	From another point of view, from the expression of $\sum_{i=1}^{N} {\mathbb{E}_{t_i}} \{u_i(t_i)\}$ in the proof of Proposition~\ref{thm:Prop.Seller} (\cite{Yang.LongVersion.2013}), the user's optimal utility after the allocation can be expressed as follows:
		\begin{equation} \label{fml:UserUtilAltExpr}
			u_i^*(t_i) = {\mathbb{E}_\mathbf{t}} \Big\{ \sum\limits_{j=1}^{M}\frac{1-F(t_i)}{f(t_i)}\frac{\partial{v_{ij}(t_i)}}{\partial{t_i}} p_{ij}^*(t_i) \Big\}.
		\end{equation}
	As we have assumed, the valuation functions of users are non-decreasing and concave. Therefore, $\frac{\partial{v_{ij}(t_i)}}{\partial{t_i}}$ decreases with respect to $t_i$. Then, the utility is affected by $\frac{1-F(t_i)}{f(t_i)}$ which is the distribution pattern of user's type and $p_{ij}^*(t_i)$ which is the amount of resources allocated to the user.

	For the service provider, the optimal revenue from selling resources can be calculated by placing the optimal allocation $p^*$ into the revenue function defined in (\ref{fml:SellerMaxObj}). According to the second term in~(\ref{fml:SellerMaxObj}), the service provider's revenue is also affected by the marginal utility of the users. Moreover, the premium and discount term (i.e., $S_i(t_i)$) affects the optimal revenue, depending on users' choices of coefficients $h_i$ and $l_i$ (see (\ref{fml:STerm})).
	
	In the following section, we will use examples and numerical results to support the analyses of utility and revenue functions.

%--------------------------------------------------------------------
\section{Example Scenarios and Numerical Results} \label{tag:sec.numerical}
In this section, we present examples and numerical simulation results of the proposed auction model for the substitutable and complementary resources. We discuss a model with two mobile cloud users for simplicity. The mobile cloud service provides two servers for computational tasks, as well as two communication channels to support different transmission rate (or bandwidth) requirements. We consider a symmetric and an asymmetric scenarios. Then, we show the impact caused by the premium and discount factors to the user's utility and service provider's revenue.

Suppose there are four commodities to be auctioned belonging to the service provider. Two of the resources are transmission channels for the buyers to choose (i.e., transmission capacity) indexed as $j=1$ and $j=2$. The other two resources are servers (i.e., commoditized computational capacity), indexed as $j=3$ and $j=4$. Therefore, there are two complementary groups in the service provider, each of the groups contains two substitutable resources, i.e., $G=2$, and $M_1=M_2=2$ as shown in Fig.~\ref{fig:SystemModel}(b).

	\subsection{Example 1: A Basic Case}
	First we consider a basic case where buyers' preferences on either complementary or substitutable resources are zero. That is, $\forall{i}, l_i(t_i) \equiv h_i(t_i) \equiv 0$. The optimal mechanism $(p^*,c^*)$ as in (\ref{fml:SellerOptCost}) and (\ref{fml:SellerMaxObj}) reduces to
		\begin{equation*}
			\begin{aligned}
			p^* & = \mathop{\arg\max}\limits_{p} \sum\limits_{i=1}^{2}{\mathbb{E}_{\mathbf{t}}} \Big\{ \sum\limits_{j=1}^{4} \big( v_{ij}(t_i) - \frac{1-F(t_i)}{f(t_i)}\frac{\partial{v_{ij}(t_i)}}{\partial{t_i}} \big)p_{ij}(t_i)\Big\} \\
			c_i^*(t_i) & = {\mathbb{E}_\mathbf{t}} \Big\{ \sum\limits_{j=1}^{4}p_{ij}^*(t_i)v_{ij}(t_i) - \sum\limits_{j=1}^{4}\int\limits_{\underline{t_i}}^{t_i}\frac{\partial{v_{ij}(x)}}{\partial{t_i}}p_{ij}^*(x)dx \Big\}.
			\end{aligned}
		\end{equation*}
		subject to (\ref{fml:Rational}), (\ref{fml:IncentiveUIntg}) and (\ref{fml:IncentiveFinalCond}). As a result, the auction becomes a general single-commodity auction case as in \cite{Myerson.OAD.1981}.

	\subsection{Example 2: A Symmetric Case}
	We consider a two-buyer, symmetric scenario as follows:
		\begin{itemize}
			\item In this symmetric case, we discuss the situation that the buyers' valuations on the first group of resources ($j=1,2$) are linear, i.e., $v_{i1}=\gamma_{i1}t_i$ and $v_{i2}=\gamma_{i2}t_i$. The expressions of $v_{i1}$ and $v_{i2}$ mean that user $i$'s valuation on a communication channel (resources 1 and 2) is linear with respect to the type. The relation between valuation functions and types indicates that the valuation linearly increases as the buyers' preference to the resources (transmission bandwidth) increases.
			
			\item The buyers' valuations on the second group of resources ($j=3,4$) are increasing and concave functions of types ($\frac{\partial f(x)}{\partial x}\geq0$ and $\frac{\partial^2f(x)}{\partial x^2}\leq0$), where the marginal payoffs decrease as the types increase. We consider $v_{i3}=\log(1+\theta_{i3}/t_i)$ and $v_{i4}=\log(1+\theta_{i4}/t_i)$, $i=3,4$. Similarly, the type $t_i$ in $v_{i3}$ and $v_{i4}$ represents user $i$'s preference on computational capacity (i.e., servers). The intuition behind such concave valuations is that the corresponding resources might not be the main bottleneck of performance. As a result, the buyers still prefer higher level resources but with less desire.
		
			\item User's type $t_i$ is uniformly distributed from 0 to 1, for $F(t_i)=t_i$ and $f(t_i)=1$. Also, the lower bound of $t_i$ is $\underline{t_i}=0$ and upper bound of $t_i$ is $\overline{t_i}=1$.
		\end{itemize}
		
		The optimal resource allocation $p^*$ as in (\ref{fml:SellerMaxObj}) is optimized as follows:
		\begin{equation} \label{fml:SymmAllc}
			\begin{aligned}
			p^* = \mathop{\arg\max}\limits_{p} & \mathbb{E}_{\mathbf{t}}\Big\{ \sum\limits_{i=1}^{2}\sum\limits_{j=1}^{4}\big[ v_{ij}(t_i)+(t_i-1)\frac{\partial v_{ij}(t_i)}{\partial t_i} \big]p_{ij}(t_i) \\
				& + \sum\limits_{i=1}^{2}h_i\big[p_{i1}(t_i)+p_{i2}(t_i)\big]\big[p_{i3}(t_i)+p_{i4}(t_i)\big] - \sum\limits_{i=1}^{2}\sum\limits_{j=1}^{4}l_i(t_i)p_{ij}(t_i) \Big\}	.
			\end{aligned}
		\end{equation}
		From the allocation scheme defined in (\ref{fml:SymmAllc}), we can see that the service provider tends to adopt combinatorial allocation as the dominant strategy. That is, to maximize the revenue, the service provider will allocate the whole group of resources to a buyer, and another group of resources as a whole to another buyer.

	\subsection{Example 3: An Asymmetric Case} \label{tag:subsec.asym}
	A simple asymmetric auction scenario is considered. The basic parameter setting is the same as that of the symmetric scenario, except the users' valuation functions are heterogeneous, as follows:
	\begin{itemize}
		\item The valuation functions of user 1 are strictly concave. We assume in this case that $v_{11}=\gamma_{11}\sqrt{t_1}$, $v_{12}=\gamma_{12}\sqrt{t_1}$, $v_{13}=\sqrt{t_1}\log(1+\theta_{13}/\sqrt{t_1})$ and $v_{14}=\sqrt{t_1}\log(1+\theta_{14}/\sqrt{t_1})$.
		\item The valuation functions of user 2 are different from that of user 1, i.e., $v_{21}=\gamma_{21}{t_2}$, $v_{22}=\gamma_{22}{t_2}$, $v_{23}={t_2}\log(1+\theta_{23}/{t_2})$ and $v_{24}={t_2}\log(1+\theta{24}_b/{t_2})$. We set in our examples that $\forall{i}\in\{1,2\}$, $\gamma_{i1}=10$, $\gamma_{i2}=20$, $\theta_{i3}=1.25\times{10}^6$ and $\theta_{i4}=3.75\times{10}^6$.
		\item The distribution of type $t_1$ is not uniform, i.e., $F(t_1)=(t_1)^2$ and $f(t_1)=2t_1$, where $t_1\in[0,1]$, while type $t_2$'s distribution is still uniform over $[0,1]$.
	\end{itemize}

\begin{figure*}[!t]
		\begin{center}
		\subfloat[]{\includegraphics[scale=0.39]{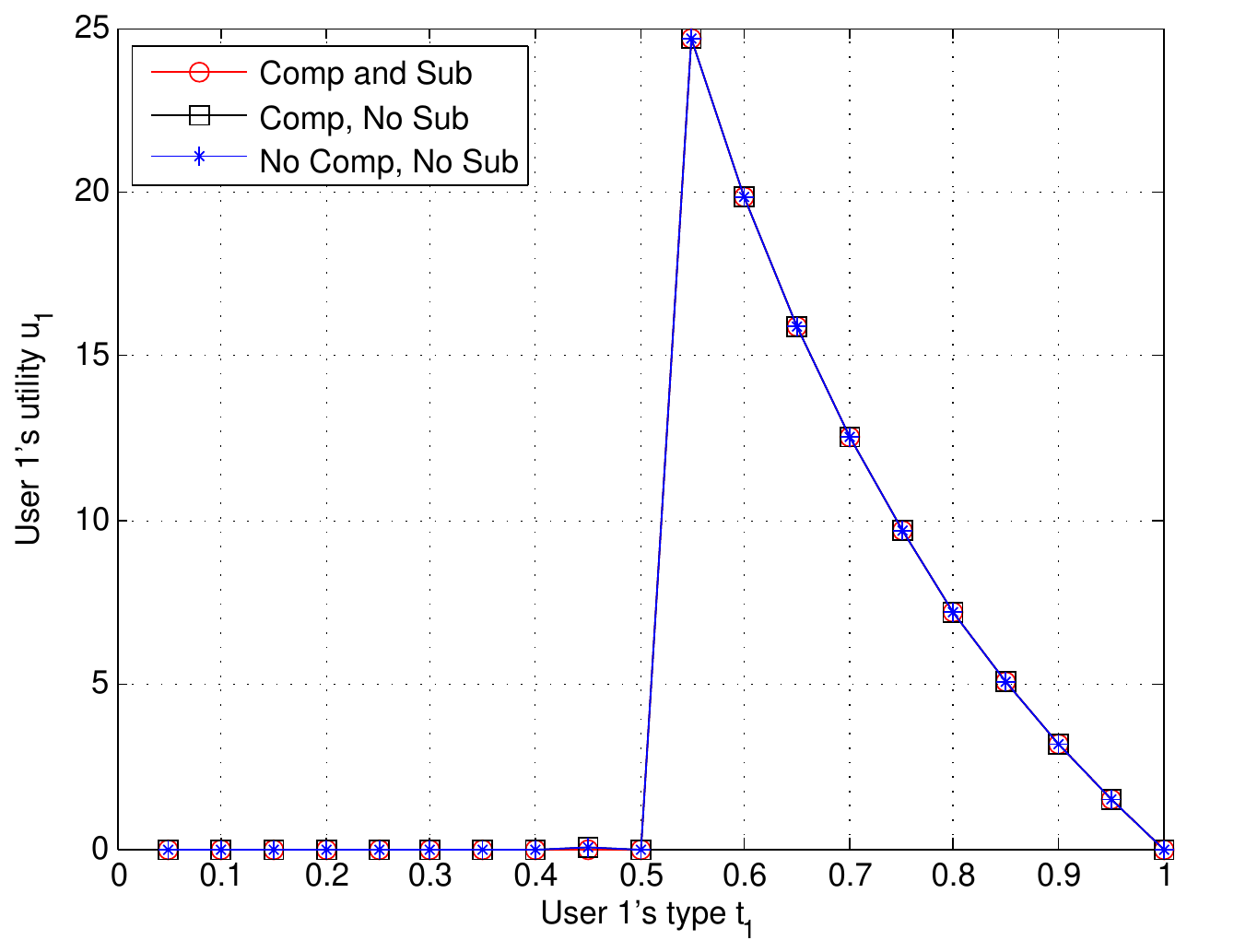}}
		\subfloat[]{\includegraphics[scale=0.39]{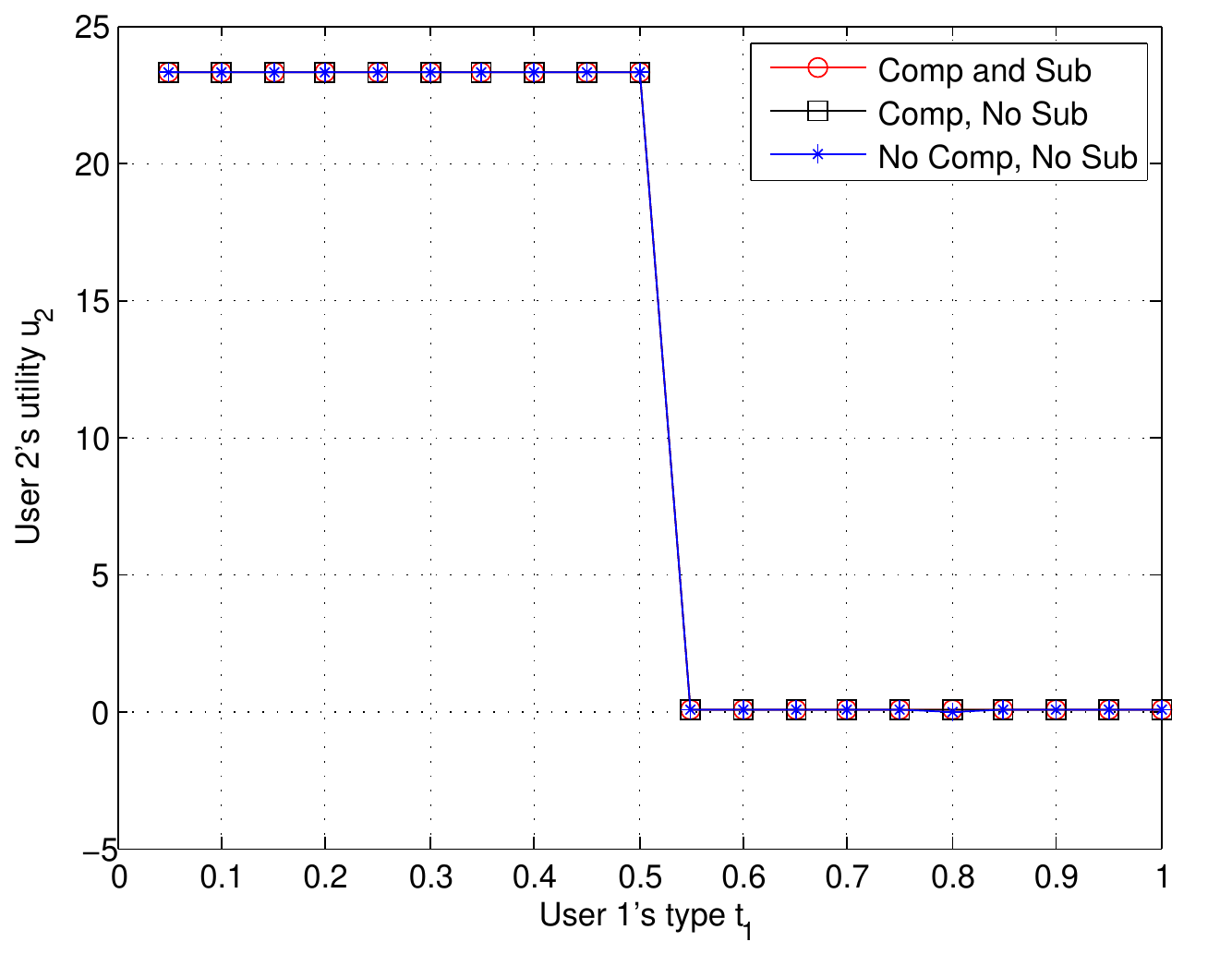}} \\
		\subfloat[]{\includegraphics[scale=0.39]{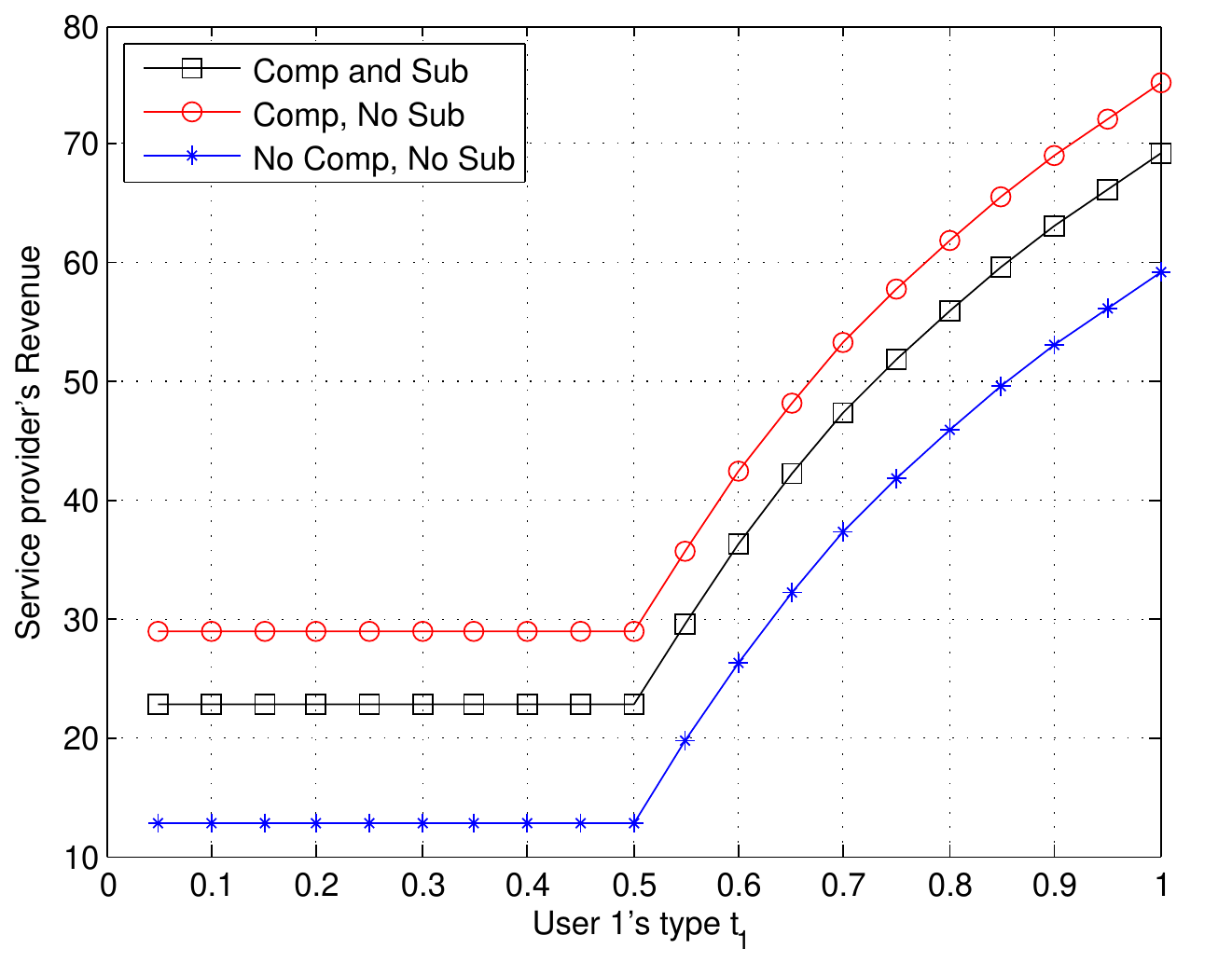}}
		\subfloat[]{\includegraphics[scale=0.39]{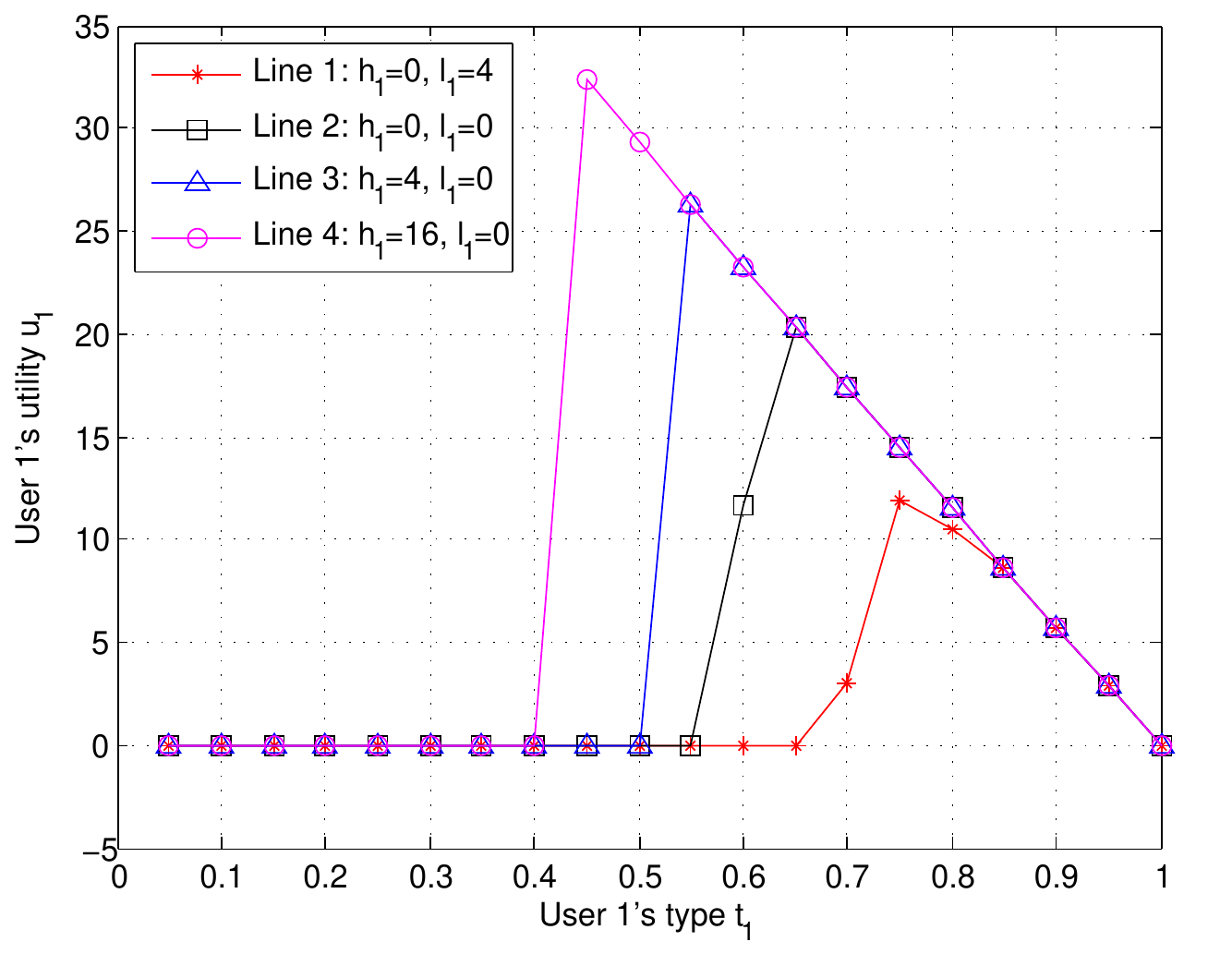}}
		\end{center}
		\caption{\small Asymmetric scenario: (a) user 1's utility to her type $t_1$, (b) user 2's utility to user 1's type $t_1$, and (c) the service provider's revenue to user 1's type $t_1$. Premium and discount factors' impacts: (d) user's utility under different settings of premium and discount factors.}
		\label{fig:NumAsym}
		\end{figure*}

	We fix user 2's type $t_2$ at 0.6, and gradually increase user 1's type $t_1$ from 0 to 1. As shown in Figs.~\ref{fig:NumAsym}(a) and (b), when $t_1$ is small, user 2 is the winner and contributes a flat revenue to the service provider. When $t_1$ increases, user 1's valuation may surpass that of user 2. User 1 becomes the winner after a turning point (0.5 in this case). The decreasing of marginal utility causes the user 1's utility to decrease as in Fig.~\ref{fig:NumAsym}(a). Also, derived from (\ref{fml:UserUtilAltExpr}), the utility of user 1 is convex after the turning point as shown in Fig.~\ref{fig:NumAsym}(a).

		%
	
	%---------------------------------------
	%---------------------------------------
	\subsection{Impacts of Premium and Discount Factors}
	Fig.~\ref{fig:NumAsym}(c) resulted from the asymmetric scenario (Section \ref{tag:subsec.asym}) depicts the effects of complementary and substitutable resources to the revenue of the cloud service provider (i.e., seller). It is shown that the buyers will have extra utility on the complements, so that the service provider's revenue will increase according to (\ref{fml:SellerMaxObj}). Also, the existence of substitutable resources will decrease the total revenue.
	
	In the examples, the premium and discount factors of users can be set in 3 different cases as follows:
	\begin{itemize}
		\item With both complementary and substitutable resources: $h_1=h_2=4$ and $l_1=l_2=1.5$. In this case, the resources in the same group are substitutable for the users with a discount factor of 1.5, and resources from different groups are complementary with a premium factor of 4.
		\item With complementary resources only: $h_1=h_2=4$ and $l_1=l_2=0$. The resources from different groups are complementary. However, the resources from the same group are not substitutable.
		\item Without any complementary and substitutable resources: $h_1=h_2=0$ and $l_1=l_2=0$. 
	\end{itemize}
	
	From Fig.~\ref{fig:NumAsym}(d) we can compare the auction results given the different settings of premium and discount factors. It is clear that the system with only complementary resources has the optimal performance in terms of utilities of users and revenue of the service provider, compared with other cases when other settings are the same. This result is intuitive, since the premium factor of the complementary resources increases the revenue of the service provider according to (\ref{fml:SellerMaxObj}). However, for the system with both complementary and substitutable resources, and the system without any complementary and substitutable resources, the results may vary, since the discount factor might degrade the performance. 

	For a user, the premium and discount factors can affect the resource allocation done by the service provider. According to (\ref{fml:SellerMaxObj}), the premium factor increases the total revenue, so the service provider is in favor of allocating resources to the user with a higher premium factor. For the discount factor, the effect is just opposite. Fig.~\ref{fig:NumAsym}(d) shows the impact of premium and discount factors to the user 1's utility. In this case, user 2's type $t_2$ is fixed at 0.6, and has the premium and discount factors set as $h_2=l_2=0$. Then, $t_1$ increases from 0 to 1. The parameter setting is the same as that used in the symmetric scenario except $h_i$ and $l_i$. Line 2 (see legend of Fig.~\ref{fig:NumAsym}(d)) is a reference result for $h_1=l_1=0$. It is shown by the figure that, a higher premium factor will increase the user's utility, and thus the total revenue is increased. The service provider may allocate the resources to the user even the type of the user is relatively low. Similarly, the existence of high discount factor will decrease the user's chance of being allocated with the resources. 

%--------------------------------------------------------------------
\section{Summary} \label{tag:sec.conclusion}
In this paper, we have addressed an analytical auction model for the resource allocation problem in a mobile cloud computing system. The complementary and substitutable cloud resources have been considered. We have analyzed the model, and solved the optimization problem to maximize the revenue of the service provider given the proposed auction mechanism. From the numerical results, we have found the changing of users' utilities and the service provider's revenue with respect to users' types, the optimal allocation schemes of the service provider, and the impacts of users' premium and discount factors to the users' utilities and the service provider's revenue.

%--------------------------------------------------------------------
% The following appendix will only be appeared in the online extended version of this paper
% The following appendix will only be appeared in the online extended version of this paper
% The following appendix will only be appeared in the online extended version of this paper
%--------------------------------------------------------------------
\section*{Appendix} \label{tag:sec.appendix}
Proof of Proposition 1:

	\begin{proof}
	First, we show the situation when the mechanism $(p,c)$ is individually rational and incentive compatible. Replace $t_i$ with $\hat{t}_i$ in (\ref{fml:RealTypeUtil}) (i.e., we treat $\hat{t}_i$ as the true type, and $v_{ij}(\hat{t}_i)$ as the user's true valuation), i.e., 
		\begin{equation}
			u_i(\hat{t}_i) = {\mathbb{E}_{\mathbf{t}_{-i}}}\Big\{ \sum\limits_{j=1}^{M}p_{ij}(\hat{t}_i)v_{ij}(\hat{t}_i) + S_i(\hat{t}_i) \Big\} - c_i(\hat{t}_i) .
		\end{equation}
		
	The incentive compatibility condition in (\ref{fml:Feasible}) can be thus written as follows:
		\begin{equation} \label{fml:IncentiveGeq1}
			u_i(t_i) \geq u_i(\hat{t}_i) + {\mathbb{E}_{\mathbf{t}_{-i}}}\Big\{ \sum\limits_{j=1}^{M}\big{(}v_{ij}(t_i)-v_{ij}(\hat{t}_i)\big{)}p_{ij}(\hat{t}_i) \Big\}.
		\end{equation}
	
	As the term $S_i(t_i,\mathbf{t}_{-i})$ and $c_i(t_i)$ are canceled from both sides of (\ref{fml:IncentiveGeq1}), the rest part of proof is similar to the procedure in \cite{Branco.MUAI.1996}. We obtain
		\begin{equation} \label{fml:IncentiveUDiff}
			\frac{\partial{}}{\partial{t_i}}u_i(x) = {\mathbb{E}_{\mathbf{t}_{-i}}}\Big\{ \sum\limits_{j=1}^{M}\frac{\partial{v_{ij}(x)}}{\partial{t_i}} p_{ij}(x) \Big\}
		\end{equation}
	whose integral form can be expressed as in (\ref{fml:IncentiveUIntg}). Meanwhile, (\ref{fml:IncentiveGeq1}) and (\ref{fml:IncentiveUIntg}) imply (\ref{fml:IncentiveFinalCond}). Finally, (\ref{fml:Rational}) intuitively holds based on the individual rationality precondition.

	Secondly, when the conditions in (\ref{fml:Rational}) to (\ref{fml:IncentiveFinalCond}) hold, the individual rationality property can be proved by (\ref{fml:IncentiveUIntg}) and (\ref{fml:IncentiveFinalCond}) since we have assumed earlier that $v_{ij}\geq0$ and it is a concave function to $(t_i)$. By replacing $\underline{t_i}$ in (\ref{fml:IncentiveUIntg}) with any fake type $\hat{t}_i\in[\underline{t_i},\overline{t_i}]$ (i.e., the user does not submit a real valuation), the incentive compatibility property is proved.
	\qed
	\end{proof}
	
Proof of Proposition 2:

	\begin{proof}
	According to (\ref{fml:RealTypeUtil}) and (\ref{fml:STerm}), we have the revenue of the service provider expressed as follows:
		\begin{equation} \label{fml:OptCost}
			\sum\limits_{i=1}^{N}{\mathbb{E}_{t_i}}\Big\{c_i(t_i)\Big\} = \sum\limits_{i=1}^{N}{\mathbb{E}_\mathbf{t}} \Big\{ \sum\limits_{j=1}^{M}p_{ij}(t_i)v_{ij}(t_i) + S_i(t_i) \Big\} - {\mathbb{E}_{t_i}}\Big\{\sum\limits_{i=1}^{N}u_i(t_i)\Big\}
		\end{equation}
	where
		\begin{equation} \label{fml:OptCost.Term3}
			\sum\limits_{i=1}^{N} {\mathbb{E}_{t_i}} \{u_i(t_i)\} = \sum\limits_{i=1}^{N} \Big\{ u_i(\underline{t_i}) + {\mathbb{E}_\mathbf{t}} \sum\limits_{j=1}^{M} \frac{1-F(t_i)}{f(t_i)}\frac{\partial{v_{ij}(t_i)}}{\partial{t_i}}p_{ij}(t_i) \Big\}.
		\end{equation}
		
	The result in (\ref{fml:OptCost.Term3}) is derived from (\ref{fml:IncentiveUIntg}), as well as the derivation process~(11) in \cite{Branco.MUAI.1996}. The integral term in (\ref{fml:IncentiveUIntg}) is transformed into a simple fraction term, if the CDF and PDF of each user $i$'s type are known. After the transformation, the expression of the utility is simplified.
	
	Based on~(\ref{fml:OptCost}) and~(\ref{fml:OptCost.Term3}), the expression of the service provider's total revenue is,
		\begin{equation} \label{fml:SellerRevenueU}
			\sum\limits_{i=1}^{N}{\mathbb{E}_\mathbf{t}} \Big\{ \sum\limits_{j=1}^{M} \big( v_{ij}(t_i) - \frac{1-F(t_i)}{f(t_i)}\frac{\partial{v_{ij}(t_i)}}{\partial{t_i}} \big)p_{ij}(t_i) + S_i(t_i) \Big\} - \sum\limits_{i=1}^{N}u_i(p_{i\cdot},c_i,\underline{t_i})
		\end{equation}
	where variable $c=\{c_1,c_2,\ldots,c_N\}$ only appears in the last term of (\ref{fml:SellerRevenueU}), i.e., $u_i(p_{i\cdot},c_i,\underline{t_i})$. As a result, to maximize $\sum_{i=1}^{N} {\mathbb{E}_{t_i}} \{ c_i(t_i) \}$, the service provider controls the cost $c$, such that the term \begin{footnotesize}$u_i(p_{i\cdot},c_i,\underline{t_i})$\end{footnotesize} is zero which is the smallest possible value (see (\ref{fml:Rational})). If $\sum\nolimits_{i=1}^{N}u_i(p_{i\cdot},c_i,\underline{t_i})=0$, (\ref{fml:SellerRevenueU}) becomes the objective function defined in (\ref{fml:SellerMaxObj}). Thus we can solve for $p^*$ from the maximization problem (\ref{fml:SellerMaxObj}). Since the total revenue (\ref{fml:SellerMaxObj}) is the sum of cost charged to each user, we also obtain the the optimal charged cost to user $i$ as in (\ref{fml:SellerOptCost}).
	\qed
	\end{proof} 


% \clearpage
% \addtocmark[2]{Author Index} % additional numbered TOC entry
% \renewcommand{\indexname}{Author Index}
% \printindex
% \clearpage
% \addtocmark[2]{Subject Index} % additional numbered TOC entry
% \markboth{Subject Index}{Subject Index}
% \renewcommand{\indexname}{Subject Index}
% \input{subjidx.ind}
\end{document}